\documentclass[a4paper,twoside]{article}
\newcommand{\affil}[1]{$^{\rm #1}$}
\usepackage[authoryear]{natbib}
\bibpunct{(}{)}{;}{a}{}{,}
\usepackage{graphicx}
\date{} 
\title{\large\bf\flushleft The $h$-index in Australian Astronomy}
\author{\parbox{\textwidth}{\flushleft
\vspace{-0.5cm}
{\it Kevin A.\ Pimbblet\affil{A,B} }\\
\vspace{0.4cm}
{\small \affil{A}\,School of Physics, Monash University, Clayton, VIC 3800, Australia}\\
{\small \affil{B}\,Email: Kevin.Pimbblet@monash.edu}}}
\begin{document}
\twocolumn[
\begin{changemargin}{.8cm}{.5cm}
\begin{minipage}{.9\textwidth}
\vspace{-1cm}
\maketitle
%
%
\small{\bf Abstract:}
The Hirsch (2005) $h$-index is now widely used as a metric
to compare individual researchers.  To evaluate it in the
context of Australian Astronomy, the $h$-index for every
member of the Astronomical Society of Australia (ASA) is found
using NASA's Astrophysics Data System Bibliographic 
Services (ADS).
Percentiles of the $h$-index distribution are detailed for
a variety of categories of ASA members, including students.
This enables a list of the top ten Australian researchers 
by $h$-index to be produced.  These top
researchers have $h$-index values in the range $53<h<77$,
which is less than that recently 
reported for the American Astronomical
Society Membership.  We suggest that membership of extremely 
large consortia such as SDSS may partially explain the difference.
We further suggest that many student ASA
members with large $h$-index values have probably
already received their Ph.D.'s and need to upgrade their
ASA membership status. 
To attempt to specify the $h$-index distribution relative
to opportunity, we also detail the percentiles of
its distribution by years since Ph.D.\ award date.  
This shows a steady increase in $h$-index with seniority, 
as can be expected.

\medskip{\bf Keywords:} sociology of astronomy ---
publications, bibliography ---
astronomical data bases: miscellaneous


\medskip
\medskip
\end{minipage}
\end{changemargin}
]
\small

\section{Introduction}

The modern research academic is judged as never before: a large variety of
metrics are now employed to determine the worth and merit of researchers,
particularly when it comes to hiring.  Anecdotally, one of the chief metrics used
is the Hirsch index ($h$-index; Hirsch 2005).
The $h$-index is formally defined as follows:
``A scientist has index $h$ if $h$ of 
his or her $N_p$ papers have at least 
$h$ citations each and the other 
($N_p - h$) papers have $\leq h$ citations each'' (Hirsch 2005).
Its modest
simplicity is probably a prime factor in its rapid pick-up
by major publishers (Ball 2005; Anon 2005).
Moreover, this index is particularly useful as it has superior predictive power 
(in terms of productivity) for the
future of researchers compared to the total number of career citations, career 
publications and mean citations per paper (Hirsch 2007).
Although other metrics and analyses exist (cf.\ Pearce 2004; Kurtz et al.\ 2005;
Egghe 2006;
Kosmulski 2006;
Jin 2006;
Blustin 2007; 
Jin et al.\ 2007;
Bornmann, Mutz \& Daniel 2008;
Wu 2010; Zyczkowski 2010), the $h$-index remains as the most prominent
of its class in the field.

Recently, Conti et al.\ (2011) presented work on the Astronomer's H-R Diagram
(number of Google search results versus citations and $h$-index) for members
of the American Astronomical Society (AAS).  Contained within that presentation
are a number interesting concepts: a top-ten list of $h$-index of AAS members
(spanning the range $94<h<118$) and the $h$ indices of all AAS members.
This work is motivated by the Conti et al.\ (2011) presentation and seeks
to determine the typical range of $h$-index in Australian Astronomy which 
may be of use for future employers and employees in the community.
The format of this work is as follows.
In Section~2, we give an overview of the dataset that we use: the membership
of the Astronomical Society of Australia.
In Section~3 we determine percentiles of the $h$-index distribution
for a variety of ASA membership categories, including students. 
To attempt to normalize relative to opportunity, we 
re-evaluate the $h$-index distribution as a function 
time elapsed since Ph.D.\ award date in Section~4.  Our conclusions are
presented in Section~5.

\section{Data}
To determine the $h$-index of Australian astronomers, we make use of the
Astronomical Society of Australia (ASA) membership list.  The membership
list is a fair representation of the Australian astronomical community: the majority
of professional astronomers are members.
Membership of the ASA comes in several difference categories which we 
use a single letter to abbreviate and detail in
Table~\ref{tab:cats}.  The advantage of the ASA membership list is that 
we can differentiate different grades of members (i.e.\ amateurs from 
professional astronomers who actively publish) to better probe
the $h$-index in these sub-categories.

\begin{table*}
\begin{center}
\caption{Categories of ASA membership, showing the number of members (N) in each
category and the fraction of each that were excluded (X) from the subsequent analysis
due to name confusion (see Section~2).
We ignore the extra categories of
Associate Society member (i.e.\ persons who are members of other learned societies that
are likely to not possess Ph.D.'s in astronomy) and corporate members.
It is important to note that individual researchers can belong
to multiple categories; e.g.\ retired, overseas fellows. 
}\label{tab:cats}
\begin{tabular}{llll}
\hline Category & N & X & Notes \\
\hline 
M & 262 & 0.10 & Member: Full professional member with a Ph.D.\ in astronomy or related discipline. \\
F & 81 & 0.10 & Fellow: Senior members of the community with potentially decades of experience. \\
S & 164 & 0.09 & Student: Post-graduate students studying toward Ph.D.'s in astronomy. \\
H & 15 & 0.00 & Honourary: Elected by the ASA council for distinguished contributions. \\
R & 54 & 0.11 & Retired.\\
O & 56 & 0.10 & Overseas.\\
A & 14 & 0.00 & Associate Members: \\
  & & & Educators, communicators and amateur astronomers lacking a Ph.D.\\
\hline
\end{tabular}
\medskip\\
\end{center}
\end{table*}

For each ASA member, we then implement a search in NASA's
Astrophysics Data System (ADS) to return a list of all refereed publications.
We then sort this list according to citations to determine the $h$-index for
each ASA member.  We note for prosperity that these searches were implemented
on 24th--25th January 2011 and were correct on a best-efforts basis as of said 
date range.

A big issue in this methodology is attempting to tie down each individual to
unique entries in ADS.
Although the present author is blessed with a very rare surname, 
others in the community are not.  For more common surnames, we use the
first name and the middle initials to help determine the $h$-index
of specific researchers, including attempting common substitutions 
for first names; e.g., `Bill' for
`William', etc. 
However, for the very common surnames (e.g.\ Smith),
this is not always possible.  Therefore, the subsequent analysis in this
work does not include any names for which we could not adequately 
differentiate a single individual in the literature in a reasonable amount
of time.  
This affects $\sim$10\% of the membership list and is labelled 
X in Table~\ref{tab:cats}.  
We caution that
the subsequent analysis should therefore be regarded as incomplete:
the inclusion of these names could increase or decrease the relative
rankings of individuals within the ASA membership.
We also note that we make no attempt to exclude self-citations
in our analysis (e.g.\ Pimbblet 2011).  Finally, it may be
the case that some of the categories may not be up-to-date due 
to (e.g.) student members gaining their Ph.D.'s and either not
upgrading to full membership status immediately or the list itself not
being updated immediately.  

\begin{figure}  
\begin{center}
\includegraphics[scale=1, angle=0, width=3.4in]{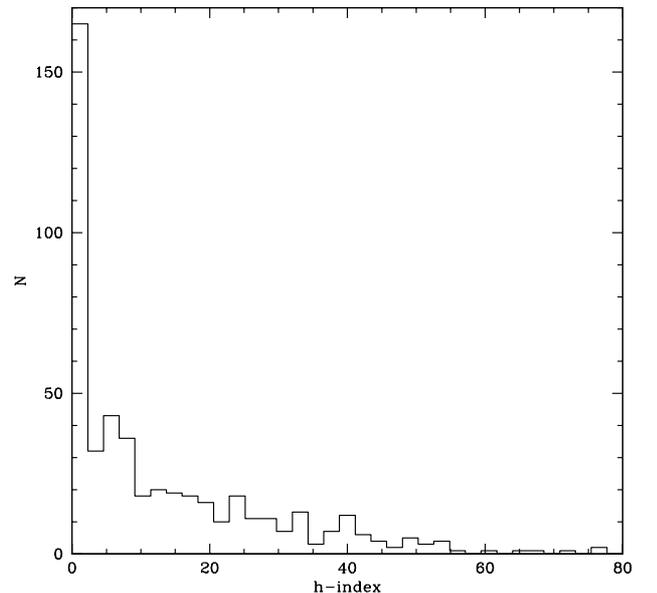}
\caption{Histogram of $h$-index for all ASA members, excluding
those for whom name confusion could not be resolved.}\label{fig:hist}
\end{center}
\end{figure}

\section{$h$-index by ASA Membership Category}
This simplest point of departure for the $h$-index analysis is to
pull out the top ten: those people who could rightly be called academic
giants in their own right in the community.
To do this, we simply rank all professional
members who are not based
overseas (i.e.\ categories M+F+S+H+R; Table~\ref{tab:cats}).
This top ten list is presented in Table~\ref{tab:topten}.

Although we refrain from commenting on individuals in this
list, it is instructive to compare it to Conti et al.'s (2011)
list.  The $h$-index values for the top ten AAS members is much
higher than the Australian ($94 < h < 118$ versus $53 < h < 77$).
Examination of Conti et al.'s (2011) figures suggests that membership
of very large observational programmes such as the Sloan Digital Sky
Survey (SDSS; e.g.\ Abazajian et al.\ 2009) can boost researcher's
$h$-index above mean values.  It is certainly the case that the
Australian top ten is dominated by non-SDSS professionals and 
we therefore suggest that most of the difference seen between
the two samples could be due to this effect. Indeed, seven of
the AAS's top ten (Conti et al.\ 2011) 
are contained in the author list of Abazajian et al.\ 
(2009). However, we do note that the Australian top ten does
contain a number of members of other consortia (not as large or
extensive as SDSS) such
as the 2dF Galaxy Redshift Survey (e.g.\ Colless et al.\ 2001).
Moreover, the majority of the listed researchers 
in Table~\ref{tab:topten} also feature 
in Thomson Reuters' ISI's highly cited list (isihighlycited.com)
for space science.

\begin{table}[h]
\begin{center}
\caption{Top 10 $h$-index for Australian Astronomy, excluding 
overseas professionals.}\label{tab:topten}
\begin{tabular}{lll}
\hline Rank & Name & $h$ \\
\hline 
=1 & Ken FREEMAN & 77 \\
=1 & Jeremy MOULD & 77 \\
3 & Karl GLAZEBROOK & 71 \\
4 & Dick MANCHESTER & 68 \\
5 & Michael DOPITA & 64 \\
6 & Warrick COUCH & 61 \\
7 & Matthew COLLESS & 56 \\
8 & Brian SCHMIDT & 54 \\
=9 & Mike BESSELL & 53 \\
=9 & Joss BLAND-HAWTHORN & 53 \\
\hline
\end{tabular}
\medskip\\
\end{center}
\end{table}

But what of the rest of the community?  In Figure~\ref{fig:hist} we 
display a histogram of $h$-index for all ASA members.  This graph
is dominated by those members having a zero $h$-index or slightly
above, much as the AAS community is (Conti et al.\ 2011).
The vast majority of these members are student members, many of whom
are likely to not have published.  Even if they have published, the
duration of the Ph.D.\ may mean that sufficient time has not elapsed
to gain large numbers of citations and that only the very exceptional
papers produced by students garner large number of citations immediately.
Clearly students in present-day 
collaborations such as WiggleZ (Drinkwater et al.\ 2010)
will benefit from this effect much in the same way as SDSS members
receive a boost.

To analyse the content of Figure~\ref{fig:hist} in a more in-depth manner,
we now create sub-samples of the ASA membership according to grade
and determine various percentiles of the $h$-index distribution.
These percentiles are presented in Table~\ref{tab:cents}.  We do
not present results for 
individual categories H, O, R and A due to low numbers.
This can be seen in the relatively tiny difference between
the percentiles quoted for M+F+R-O versus M+F-R-O samples
in Table~\ref{tab:cents}.

We start by discussing the student membership result.  
At the upper echelons, students appear to have an $h$-index
comparable of junior professionals.  But a careful analysis
of the membership list reveals that this is exactly what these
students are: junior professionals who should be in the M
category.  
We argue that anything above the 90th percentile
for the S category should be regarded with suspicion.

Naturally, the fellows occupy much higher $h$-index values 
than the regular members do.  The effect of adding or removing the
retirees from the M+F sample is modest: the most noticeable effect
is at the upper echelons of the scale.
However, the major problem of this analysis is that it does not
specify the $h$-index relative to opportunity.  To remedy this,
we now try to divide up the ASA membership according to years
since the award of a Ph.D.  

\begin{table*}
\begin{center}
\caption{Percentiles of $h$-index distribution by ASA membership grade.
N gives the total number of members for each row.
}\label{tab:cents}
\begin{tabular}{rlllllllll}
\hline 
Category & N & \multicolumn{7}{c}{Percentile} \\
         & & 25 & 50 & 75 & 90 & 95 & 97.5 & 99 \\
\hline
M & 235 & 6.0 & 12.0 & 21.0 & 32.5 & 43.0 & 47.8 & 52.4 \\
F & 73 & 21.0 & 30.0 & 39.8 & 50.0 & 59.2 & 69.8 & 75.4 \\
S & 149 & 0.0 & 0.0 & 2.0 & 4.0 & 5.0 & 6.7 & 9.5 \\
M+F-R-O & 227 & 7.0 & 15.0 & 28.0 & 41.0 & 45.0 & 53.0 & 63.7 \\
M+F+R-O & 266 & 6.0 & 15.0 & 28.0 & 40.0 & 45.4 & 53.3 & 68.8 \\
M+F+S-R-O & 374 & 1.0 & 5.0 & 19.0 & 33.6 & 42.3 & 46.7 & 59.7 \\
\hline
\end{tabular}
\medskip\\
\end{center}
\end{table*}

\section{$h$-index by Years Since Ph.D. Award}
Even the award year of a Ph.D.\ must be regarded with healthy 
suspicion as a metric for performance relative to opportunity.
This is especially true for early-career researchers who may complete their
Ph.D.\ whilst undertaking their first post-doctoral placement
and for the many researchers who have had significant time
away from the profession; the present author included. 

To determine the award date of the Ph.D., we use the results
from ADS where available.  If the Ph.D.\ is not listed in
ADS, then we use the date of the second 
first-author refereed publication by the 
member as a compromise proxy for this date, given the
distribution of the S sample in Table~\ref{tab:cents}.  This date
was determined for all ASA members in the M+F+R-O category.
Where no date could be determined by either method, the
member was simply removed from the list.  This may 
have the effect of meaning that the percentiles for
this sample are upper limits as we have missed
doctoral researchers who have few first author publications.
We present the percentiles of this distribution in Table~\ref{tab:centsy}.
The results show a fairly steady progression as one increases 
in seniority from Ph.D.\ award date without any
obvious discrepancies, as may be expected.
However, one comment to be made is that there seems many
less young professionals in the samples than there perhaps
should be (given the numbers in more senior years).  
This tentatively suggests that new recruits to Australian
Astronomy may not be joining the ASA immediately.  

\begin{table*}
\begin{center}
\caption{Percentiles of $h$-index distribution by years since Ph.D.\ award 
(explicitly: 2011 minus the award date). 
Only members in M+R+F-O categories are included.
N gives the total number of members for each row.
}\label{tab:centsy}
\begin{tabular}{rlllllllll}
\hline
Years since & N & \multicolumn{7}{c}{Percentile} \\
Ph.D.\ award & & 25 & 50 & 75 & 90 & 95 & 97.5 & 99 \\
\hline
0--5 & 32 & 5.0 & 6.0 & 9.0 & 10.0 & 13.2 & 15.8 & 16.3 \\
6--10 & 42 & 8.5 & 14.0 & 19.0 & 22.0 & 25.0 & 27.8 & 29.7 \\
11--20 & 74 & 12.5 & 19.0 & 28.5 & 39.6 & 43.4 & 53.1 & 58.4 \\
$>$20  & 94 & 12.0 & 25.0 & 38.0 & 47.2 & 55.1 & 62.9 & 77.0 \\
\hline
\end{tabular}
\medskip\\
\end{center}
\end{table*}

Further, not all areas and sub-disciplines 
of science and astronomy may be equal.  Those researchers involved 
in (e.g.) instrumentation may have a very different $h$-index
distribution to those researching observational cosmology (particularly 
those in larger-sized consortia).

\section{Conclusions}
This work has presented an analysis of the $h$-index distributions
for present members of the ASA.  As well as deriving a top ten 
(Table~\ref{tab:topten}), we
have presented the percentiles for various sub-samples of the
ASA's membership, including student statistics (Table~\ref{tab:cents}).
We have also attempted to analyse the distribution relative to
opportunity by detailing the percentiles by time elapsed since
Ph.D.\ award date (Table~\ref{tab:centsy}). 

Clearly the $h$-index is a crude estimator of the value
of a researcher and should not be used in isolation to other
metrics, even if it is a good predictor of future 
productivity (Hirsch 2007).
It will be instructive to re-visit this analysis
in future years or decades to determine how the field 
has changed.

We terminate this work with a $caveat\ emptor$: there are known deficiencies
in this analysis such as numerous 
missing persons (who are not ASA members) 
whose statistics may alter the results presented. 
We have tried to be up-front with various caveats throughout this work, 
but there may yet be unknown unknowns present as well.
Further, there may exist transcription errors that went
un-detected during the data assembly stage.  However, as far as possible,
we believe the numbers quoted in this work are accurate.

\section*{Acknowledgments} 
KAP thanks Bryan Gaensler for tweeting about Conti et al.'s presentation
from the 2011 AAS meeting, and Michael J.\ Morgan for
many discussions on how to interpret the $h$-index 
in the context of Australian Astronomy 
which inspired this present work.  I also thank the 
anonymous referee for a positive review of the manuscript that
has improved its content.

This research has made use of NASA's Astrophysics Data System Bibliographic Services.


\end{document}